\documentclass[12pt,letter]{article}

\usepackage{graphicx, epsfig, color}
\def\eslt{E_T^{\rm miss}}
\def\to{\rightarrow}

\def\bi{\begin{itemize}}
\def\ei{\end{itemize}}
\def\te{\tilde e}
\def\DRbar{\overline{DR}}

\def\tu{\tilde u}

\def\tb{\tilde b}

\def\tst{\tilde t}
\def\ttau{\tilde \tau}

\def\tg{\tilde g}

\def\tw{\widetilde W}
\def\tz{\widetilde Z}
\def\alt{\stackrel{<}{\sim}}
\def\agt{\stackrel{>}{\sim}}
\def\be{\begin{equation}}  
\def\ee{\end{equation}}  
\newcommand\prd[3]{{\it Phys.\ Rev.\ }{\bf D #1} (#2) #3}
\newcommand\prep[3]{{\it Phys.\ Rept.\ }{\bf #1} (#2) #3}
\newcommand\prl[3]{{\it Phys.\ Rev.\ Lett.\ }{\bf #1} (#2) #3}
\newcommand\plb[3]{{\it Phys.\ Lett.\ }{\bf B #1} (#2) #3}
\newcommand\jhep[3]{{\it J. High Energy Phys.\ }{\bf #1} (#2) #3}

\newcommand\apj[3]{{\it Astrophys.\ J. }{\bf #1} (#2) #3}
\newcommand\ijmpd[3]{{\it Int.\ J.\ Mod.\ Phys.\ }{\bf D #1} (#2) #3}
\newcommand\npb[3]{{\it Nucl.\ Phys.\ }{\bf B #1} (#2) #3}
\newcommand\epjc[3]{{\it Eur.\ Phys.\ J. }{\bf C #1} (#2) #3}
\newcommand\ptp[3]{{\it Prog.\ Theor.\ Phys.\ }{\bf #1} (#2) #3}
\newcommand\zpc[3]{{\it Z.\ Physik }{\bf C #1} (#2) #3}
\newcommand{\hepph}[1]{hep-ph/#1}
\newcommand{\hepex}[1]{hep-ex/#1}
\newcommand{\astroph}[1]{astro-ph/#1}

\begin{document}
\begin{titlepage}
\begin{flushright}
FSU-HEP/080216
\end{flushright}

\vspace{0.5cm}
\begin{center}
{\Large \bf 
SUSY interpretation of the Egret GeV anomaly,\\
Xenon-10 dark matter search limits and the LHC
}\\ 
\vspace{1.2cm} \renewcommand{\thefootnote}{\fnsymbol{footnote}}
{\large Howard Baer $^{1}$\footnote[1]{Email: baer@hep.fsu.edu },
Alexander Belyaev$^2$\footnote[2]{Email: a.belyaev@phys.soton.ac.uk } and
Heaya Summy$^1$\footnote[1]{Email: heaya@hep.fsu.edu }} \\
\vspace{1.2cm} \renewcommand{\thefootnote}{\arabic{footnote}}
{\it 
1. Dept. of Physics,
Florida State University, Tallahassee, FL 32306, USA \\
2. School of Physics and Astronomy,
University of Southampton, Southampton, SO17 1BJ, UK \\
}

\end{center}

\vspace{0.5cm}
\begin{abstract}
\noindent 
The observation of the Egret experiment of an 
excess of diffuse gamma rays with energies above
$E_\gamma =1$ GeV has previously been interpreted 
in the context of the minimal supergravity model (mSUGRA) 
as coming from neutralino annihilation into mainly $b$-quarks 
in the galactic halo, with neutralino mass in the vicinity of 50-70 GeV.
We observe that in order to obtain the correct relic abundance of neutralinos
in accord with WMAP measurements, the corresponding neutralino-proton
direct detection (DD) rates should be in excess of recent limits from the Xenon-10
collaboration. 
While it does not appear possible to satisfy the Egret, WMAP and Xenon-10
constraints simultaneously within the mSUGRA model, 
we find that it is easily possible in 
models with non-universal Higgs soft masses (NUHM).
In either case, gluino pair production from $m_{\tg}\sim 400-500$ GeV should occur at 
large rates at the CERN LHC, and a gluino pair production signal should be
visible with just 0.1 fb$^{-1}$ of integrated luminosity.
The NUHM interpretation predicts a rather light spectrum of heavy Higgs bosons with 
$m_A\sim 140-200$ GeV over the whole parameter space which would interpret Egret data.
Spin-independent DD rates are predicted to be
just above $10^{-8}$ pb, within range of the next round of direct 
dark matter detection experiments.

\vspace{0.8cm}

\end{abstract}


\end{titlepage}

\section{Introduction}

An abundance of astrophysical evidence points to the conclusion
that the bulk of the matter in the universe is composed {\it not} of
Standard Model (SM) particles, but of some unknown non-relativistic 
elementary particle known as cold dark matter (CDM)\cite{cdmreview}.
An analysis of the three-year WMAP and galaxy survey data sets\cite{wmap} 
implies that the ratio of cold dark matter density to critical density, 
\be
\Omega_{CDM}h^2\equiv \rho_{CDM}/\rho_c =
0.111^{+0.011}_{-0.015}\ \ (2\sigma ) .
\label{eq:Oh2}
\ee
where $h=0.74\pm 0.03$ is the scaled Hubble constant.
While the density of CDM is becoming precisely known, the identity of 
the CDM particle (or particles) is still a complete mystery.
Although numerous candidate CDM particles populate the theoretical literature, 
the WIMPs (weakly interacting massive particles) stand out in that
their thermal abundance can be calculated, and is found to be in 
rough accord with Eq. (\ref{eq:Oh2}) provided the WIMP mass is of order
100-1000 GeV. Of the numerous WIMP candidates in the literature, 
the lightest neutralino of supersymmetric (SUSY) theories is especially popular
because SUSY solves a host of theoretical problems associated with the SM,
and also receives some (albeit indirect) support from data
(in the form of the measured gauge couplings unifying at 
$Q=M_{GUT}$ under MSSM RG evolution and also from other precision electroweak
measurements\cite{sven}).

There is at present a multi-pronged effort aimed at identifying
WIMP dark matter particles and measuring their  properties\cite{multi}. The most direct approach is to
try to detect relic WIMPs left over from the Big Bang by observing WIMP-nucleon
collisions in experiments located deep underground. Limits from the CDMS\cite{cdms} 
experiment and more recently from the Xenon-10\cite{xenon10}
experiment have begun probing the upper limits of SUSY model parameter space.

WIMP particles can also be searched for at collider experiments such as those
at the CERN LHC, especially if 
the dark matter particle is but one of a whole family of particles, some of which
can be produced via strong and electromagnetic interactions. 
The dark matter particle would then be produced by cascade decays of heavier particles, and
would lead to missing transverse energy in collider events.
Such is the case of theories such as $R$-parity conserving supersymmetry\cite{wss}, 
$KK$-parity conserving universal extra dimensions (UED)\cite{uedreview}
and little Higgs models with $T$-parity\cite{lhtreview}. 

Dark matter may also be searched for indirectly. For instance, the sun can sweep up 
WIMP particles as it traverses its galactic orbit, so that WIMPs accumulate at 
a high density in the solar core. Then WIMP-WIMP annihilation to SM particles can occur at high rates 
in the solar core. While most SM particles would be absorbed by the surrounding solar medium, 
multi-GeV scale $\nu_\mu$s would escape and later convert to muons in neutrino 
telescopes such as Amanda/IceCube or Antares. 

In addition, dark matter in the galactic halo can be
searched for indirectly via its annihilation into high energy gamma rays or anti-matter. 
In the case of gamma rays, searches look either for WIMP-WIMP annihilation directly to 
$\gamma\gamma$ pairs (loop-suppressed since WIMPs are electrically neutral) or via
WIMP-WIMP annihilation to $q\bar{q}$ pairs, followed by $q\to\pi^0\to\gamma\gamma$ via
hadronization and decay. In the latter case, one expects a diffuse spectrum of gamma rays
with energies $E_\gamma <M_{WIMP}$ atop a background arising from cosmic ray spallation
onto nuclei, inverse Compton scattering and bremsstrahlung.

The spectral shape of the gamma ray sky has been measured in the 1990s by the Egret experiment in the 
energy range of 0.1-10 GeV\cite{egret}, where already an excess of signal with $E_\gamma >1$ GeV
above expected background was noted. The $E_\gamma >1$ GeV excess apparently is seen in all
sky directions.
An analysis by de Boer {\it et al.}\cite{deboer} explains the Egret GeV anomaly as coming from
neutralino annihilation in the galactic halo into mainly $b\bar{b}$ pairs. 
A fit of the neutralino hypothesis to the Egret data favor a neutralino in the mass range
$m_{\tz_1}\sim 50-70$ GeV. 
On the astrophysics side of the de Boer {\it et al.} interpretation, 
the strength of the signal depends on the 
dark matter density distribution throughout the galaxy. In order to explain the Egret GeV anomaly, 
de Boer {\it et al.} invoke a DM density distribution involving two rings of dark matter
at 4 and 13 kpc\cite{DMdist}.

De Boer {\it et al.} further interpret the apparent WIMP annihilation signal in the context of the minimal
supergravity (mSUGRA) model\cite{deboersugra}, which allows a complete determination of all super-particle and Higgs boson
masses and mixings in terms of just a few parameters\cite{msugra}
\be
m_0,\ m_{1/2},\ A_0,\ \tan\beta\ \ {\rm and}\ \ sign(\mu) ,
\ee
where $m_0$ is a common scalar mass at energy scale $Q=M_{GUT}$, $m_{1/2}$ is the common 
gaugino mass at $M_{GUT}$, $A_0$ the common trilinear GUT-scale soft breaking mass, $\tan\beta $ is the
ratio of higgs vevs and $\mu$ is the superpotential Higgs mass parameter. The magnitude-- but not the sign--
of the $\mu$ term is determined by constraints arising from requiring an appropriate breakdown of
electroweak symmetry in the mSUGRA model. In mSUGRA, once the GUT scale soft SUSY breaking terms are
stipulated at $M_{GUT}$, their weak scale values can be calculated via renormalization group 
equations (RGEs). The physical SUSY particle masses and mixings can then be calculated
in terms of the weak scale soft SUSY breaking terms via well-known algorithms\cite{wss}.

To generate a neutralino WIMP with mass 50-70 GeV, de Boer {\it et al.} require
$m_{1/2}$ in the range of 130-170 GeV. This rather low value of $m_{1/2}$ typically leads
to light Higgs masses $m_h\alt 114.4$ GeV (the limit from LEP2 on SM-like Higgs scalars) and
to large contributions to the branching fraction $BF(b\to s\gamma )$. To avoid these constraints,
de Boer {\it et al.} adopt a large value of $m_0\sim 1500$ GeV\cite{deboer4}. Then, to gain accord with the
measured relic density (Eq. \ref{eq:Oh2}), they require large $\tan\beta\sim 54$. 
For such a large value of $\tan\beta$, the $b$-quark and $\tau$-lepton Yukawa couplings become very large,
the pseudoscalar Higgs mass $m_A$ falls, and its width grows. Neutralinos $\tz_1$ can then annihilate efficiently
in the early universe as $\tz_1\tz_1\to b\bar{b}$ via {\it virtual, non-resonant}  $A$ exchange in the
$s$-channel\cite{Aannihilation}. Just as neutralinos can annihilate efficiently via $A^*$ in the 
early universe, so can they annihilate efficiently via $A^*$ to $b\bar{b}$ in the galactic halo, 
since the $A$-annihilation diagrams are $s$-wave (while $h$, $H$ annihilation is $p$-wave, and thus
velocity-suppressed). The dominant halo annihilation $\tz_1\tz_1\to A^*\to b\bar{b}\to\pi^0\to\gamma$ 
can then describe the Egret gamma ray excess. Note that there is little uncertaintly in the 
gamma spectrum from 50-70 GeV WIMP annihilation, since the corresponding process 
$e^+e^-\to b\bar{b}\to\gamma$ has been well-measured at LEP/LEP2.

It is important to note that several alternative explanations/insights regarding the Egret 
excess have emerged\cite{hooper}.
\bi
\item In Ref. \cite{moskalenko}, Strong, Moskalenko and Reimer verify that a ``conventional model'' of
cosmic ray production and propagation is insufficient to explain the Egret GeV anomaly, even if augmented
by hard sources of additional cosmic ray production in the inner galaxy. However, by 
suitably adjusting the spectrum of cosmic protons and electrons, they find they {\it are} 
able to by-and-large match the Egret data. 
\item Alternatively, Stecker {\it et al.}\cite{stecker} 
claim the Egret excess can be explained by a calibration error in Egret gamma measurements 
with $E_\gamma >1$ GeV. This would explain why an excess of high energy gammas comes from all sky regions.
\item
Bergstrom {\it et al.}\cite{bergstrom} point out that if they use the deBoer
derived distribution of galactic dark matter, including the ring structure, then the 
SUSY region favored in the deBoer analysis yields an anti-proton flux far in excess of
measurements from the BESS experiment. These calculations adopt the isotropic DarkSUSY model of
cosmic ray propagation. DeBoer {\it et al.} counter that using an 
{\it anisotropic} model of galactic cosmic ray propagation would greatly reduce the expected $\bar{p}$
flux\cite{deboer_pbar}.
\ei

In this note, we wish to examine the SUSY interpretation of the Egret excess,
and compare it with other constraints on sparticle masses. We find that the 
SUSY interpretation in terms of the mSUGRA model is in conflict with recent results on 
direct detection of dark matter from the Xenon-10 experiment. By moving to models with non-universality
in the Higgs sector, 
however, one may preserve the SUSY interpretation of the Egret gamma ray excess, while
staying below bounds on direct detection of DM. In any case, the imminent turn-on of the LHC
should decide the issue. The deBoer interpretation predicts gluinos with mass $m_{\tg}\sim 400-500$ GeV.
Production cross sections for gluino pair production in this mass range at the LHC
are at the $10^5$ fb level. Thus, LHC should decisively test the deBoer scenario with as little as
0.1 fb$^{-1}$ of integrated luminosity. If the mSUGRA interpretation is correct, then
the heavy Higgs boson $A$ should have mass around $200-400$ GeV. 
If an interpretation in terms of the NUHM
model is correct, then $m_A$ should be much lower- in the $140-200$ GeV range.

\section{Confronting SUSY interpretation of Egret GeV anomaly with Xenon-10 direct dark matter search}

\subsection{mSUGRA analysis}

We begin by calculating sparticle mass spectra using the Isajet 7.76\cite{isajet} Isasugra code.
Isajet begins with weak scale $\DRbar$ values for the three gauge couplings and Yukawa couplings,
and evolves up in energy to determine $M_{GUT}$, defined as the $Q$ value where 
gauge couplings $g_1=g_2$. At $M_{GUT}$, soft SUSY breaking boundary conditions are input,
and the set of 26 coupled 2-loop RGEs are evolved down to $M_{weak}$. Beta-function 
threshold effects are included in the 1-loop portion of RGEs for gauge and Yukawa couplings, 
giving a smooth transition between MSSM and SM effective theories. All soft terms which mix
are frozen out at scale $M_{SUSY}=\sqrt{m_{\tst_L}m_{\tst_R}}$, while all non-mixing soft terms
are frozen out at their own mass scale ({\it e.g.} $m_{\tu_R}^2$ stops running when $Q=m_{\tu_R}$ 
is reached\cite{bfkp}. 
The RG-improved 1-loop MSSM scalar potential is minimized at $Q=M_{SUSY}$, which determines
the value of $\mu^2$. All tree level sparticle masses are computed. Once these are known, then all
1-loop sparticle masses are computed, including SUSY threshold corrections to $m_t$, $m_b$ and $m_\tau$.
The threshold effects alter the trajectories of the running couplings, so that an up-down
iterative approach is used to calculate all 1-loop corrected sparticle masses; the iterations terminate
when a convergence criterion is satisfied. At this point, all sparticle decay branching fractions
are calculated, along with neutralino relic density $\Omega_{\tz_1}h^2$, $a_\mu =(g-2)_\mu/2$,
$BF(b\to s\gamma )$, $\sigma_{SI}(\tz_1 p)$ , $BF(B_s\to \mu^+\mu^- )$ and
$\langle\sigma v\rangle |_{v\to 0}$, via the Isatools package\cite{isatools}. The latter quantity,
the neutralino annihilation cross section times relative velocity, in the limit as
$v\to 0$, is the crucial particle physics quantity needed to evaluate various halo
annihilation processes.

\begin{figure}[!t]
\begin{center}
\epsfig{file=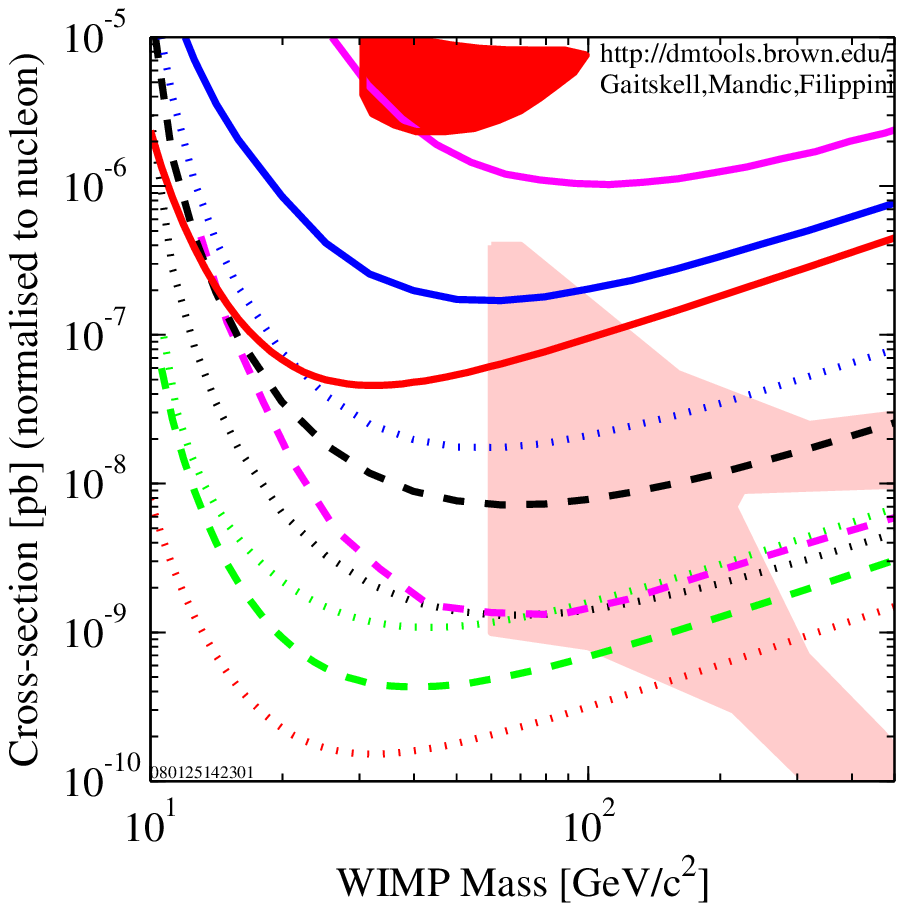,width=6cm}
\epsfig{file=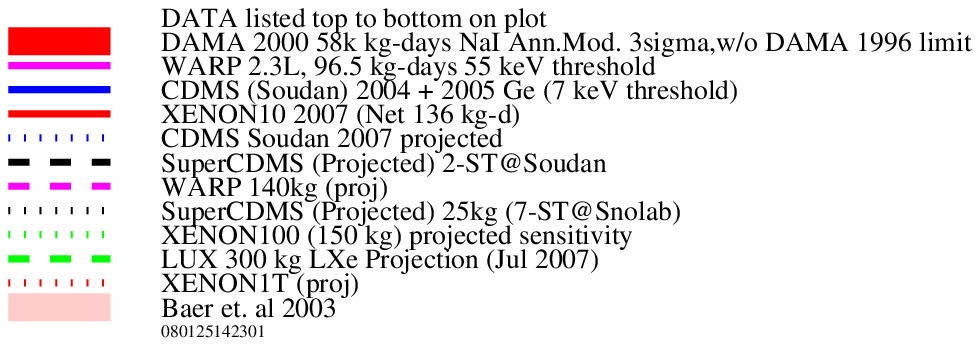,width=9cm}
\end{center}
\caption{\small\it 
Plot of reach of various present and future dark matter 
direct detection experiments compared to theory prediction
in the mSUGRA model.}
\label{fig:dd} 
\end{figure}

Regarding the neutralino direct dark matter detection cross section, 
we note here that a new limit on the spin-independent neutralino-nucleon scattering 
cross section, $\sigma_{SI}(\tz_1 p)$, has appeared from the Xenon-10 collaboration\cite{xenon10}.
This new limit, displayed as the solid red curve in Fig. \ref{fig:dd}\cite{gaitskell}, 
excludes $\sigma_{SI}(\tz_1 p)\agt 6\times 10^{-8}$ pb for $m_{\tz_1}\sim 60$ GeV. 
We also show in this figure the projected reach of several future direct detection
experiments (dashed curves), along with theoretical predictions from a scan over mSUGRA model parameter
space\cite{bbbo} (pink region). The reach contours assume a standard local DM density of
$\rho_{DM}=0.3$ GeV/cm$^3$, and a standard DM velocity profile.

\begin{figure}[!t]
\begin{center}
\epsfig{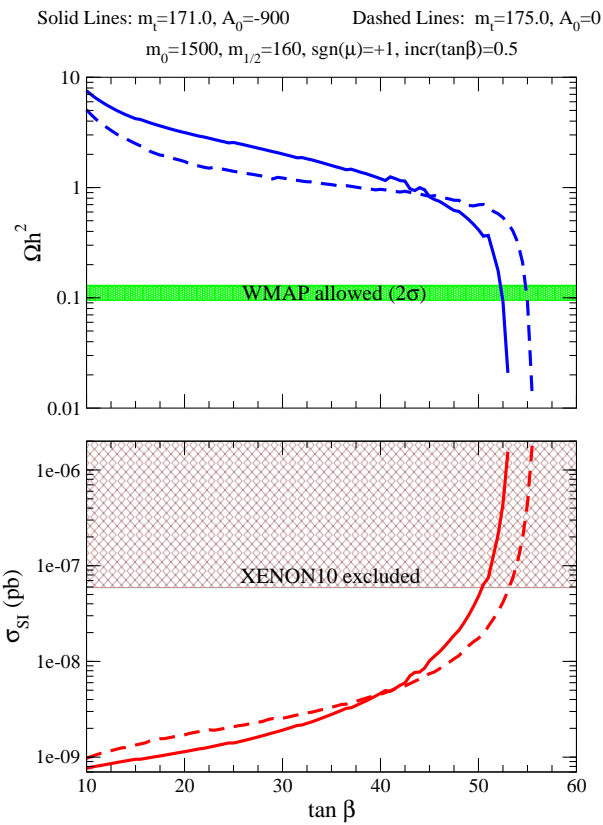}
\end{center}
\caption{\small\it 
Plot of {\it a}). $\Omega_{\tz_1}h^2$ versus $\tan\beta$ and
{\it b}). $\sigma_{SI}(\tz_1 p)$ versus $\tan\beta$ for two cases
in the mSUGRA model. }
\label{fig:eg1} 
\end{figure}

\begin{figure}[!t]
\begin{center}
\epsfig{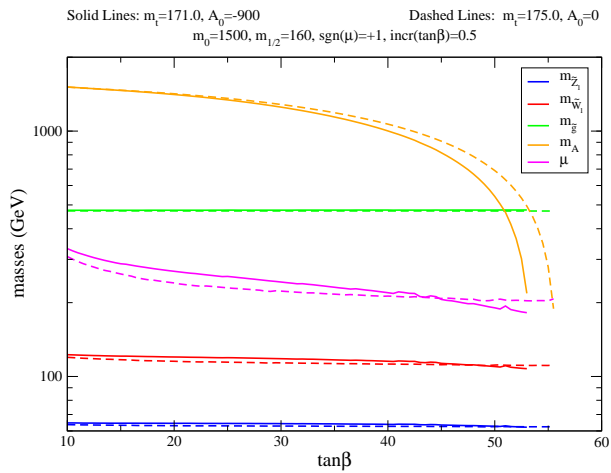}
\end{center}
\caption{\small\it 
Plot of sparticle masses versus $\tan\beta$ for the two cases
of mSUGRA model displayed in Table \ref{tab:bm}.}
\label{fig:mass} 
\end{figure}

We first attempt to verify the de Boer {\it et al.} suggested mSUGRA interpretation using the
Isajet code. The requirement that $m_{\tz_1}\sim 60$ GeV means that-- in models with gaugino mass
unification and $\mu \gg M_2$-- the lightest neutralino should be dominantly bino-like.
The trick is to get a low relic density in accord with WMAP, while maintaining a large
value of $\langle\sigma v\rangle |_{v\to 0}$, so there is sufficient neutralino  annihilation
in the galactic halo.
The LEP2 constraints that {\it i}). $m_{\tw_1}>103.5$ GeV and {\it ii}). $m_{\ttau}\agt 95$ GeV
means that co-annihilation cannot be used to reduce the relic density to WMAP allowed levels.
De Boer {\it et al.} suggest taking $m_{1/2}\sim 160$ GeV to get $m_{\tz_1}\sim 60$ GeV, and large
$m_0$ to suppress SUSY contributions to $BF(b\to s\gamma )$ and to raise the value of $m_h$
to LEP2-allowed values\footnote{Here, we require $m_h\agt 111$ GeV to account for a roughly 3 GeV
slop in the theory calculation of $m_h$, while LEP2 requires a SM-like $h$ to have 
$m_h>114.4$ GeV\cite{lep2h}.}. These input parameters, along with $A_0=0$ and $\tan\beta =10$, give $\Omega_{\tz_1}h^2\sim 10$,
which is two orders of magnitude higher than Eq. (\ref{eq:Oh2}). In Fig. \ref{fig:eg1}{\it a}., we plot the value 
of $\Omega_{\tz_1}h^2$ versus $\tan\beta$ for $m_0=1500$ GeV, $m_{1/2}=160$ GeV, $A_0=0$ and $m_t=175$
(as in de Boer {\it et al.}\cite{deboersugra}), and $m_t=171$ GeV (the central value of $m_t$ as recently measured
by D0 and CDF\cite{mtop}). While $\Omega_{\tz_1}h^2$ is too large for most of the range of
$\tan\beta$, we see that it drops to the measured value around $\tan\beta =52-55$. At this high a value
of $\tan\beta$, the $b$ and $\tau$ Yukawa couplings become very large, while the value of $m_A$
drops (see Fig. \ref{fig:mass}). Even though $2m_{\tz_1}$ is still far from the $A$ resonance, annihilation through the
virtual $A^*$ becomes dominant enough to lower the relic density to WMAP-allowed values.
The beauty of this approach is that when neutralino annihilation in the early universe via 
$s$-channel $A$ exchange is large, so also is halo annihilation of neutralinos\cite{bo}. This contrasts with the case
of $\tz_1\tz_1\to h\to b\bar{b}$, where early universe annihilation can be large, but
$\langle\sigma v\rangle |_{v\to 0}\to 0$, so that the neutralino halo annihilation rate is small.

In Fig. \ref{fig:eg1}{\it b}., we show the spin-independent neutralino-proton
scattering cross section versus $\tan\beta$ for the same parameters as in 
Fig. \ref{fig:eg1}{\it a}. We see that as $\tan\beta$ increases, the value
of $\sigma_{SI}(\tz_1 p)$ also increases. This is due to Higgs exchange direct detection
scattering diagrams, and the increasing magnitude of the $b$-quark Yukawa coupling.
The important point to notice, however, is that the value of $\sigma_{SI}(\tz_1 p)$
has increased into the Xenon-10 excluded region at a $\tan\beta$ value
somewhat lower than that needed for the relic density to enter the WMAP-allowed
dark matter band. Thus, technically, these cases for a SUSY interpretation of the
Egret GeV anomaly would be excluded by the new Xenon-10 limits.

We list two WMAP-allowed mSUGRA points in Table \ref{tab:bm}. The first, suggested in the de Boer {\it et al.}
analysis\cite{deboersugra} has $m_t=175$ GeV so that an exact comparison can be made
to Ref. \cite{deboersugra}. The second point has $m_t$ dialed down to $171$ GeV in accord with
recent top mass measurements at the Tevatron. In addition, for the second point, we take 
$A_0=-900$ GeV which raises the value of $m_h$ more closely into accord with LEP2
Higgs search limits. We see that both cases have $m_{\tz_1}\sim 60$ GeV, while
$m_{\tg}\sim 470$ GeV. Squarks and sleptons have masses above the TeV range.
We also list in the Table the relic density $\Omega_{\tz_1}h^2$, $BF(b\to s\gamma )$,
the SUSY contribution to the muon anomalous magnetic moment $\Delta a_\mu$, the
branching fraction for $B_s\to\mu^+\mu^- $ decay, $\sigma_{SI}(\tz_1 p)$ in pb,
and $\langle\sigma v\rangle |_{v\to 0}$.  
The neutralino-proton spin independent scattering cross section is $\sim 3\times 10^{-7}$ pb
in both cases-- well above the Xenon-10 limit. The value of $\langle\sigma v\rangle |_{v\to 0}\sim 2\times 10^{-26}$
cm$^3$/sec, a value which gives sufficient halo annihilation in models assuming the de Boer
DM halo profile.
%
\begin{table}
\begin{tabular}{lccc}
\hline
parameter & de Boer & mSUGRA(171) & NUHM \\
\hline
$m_0$       & 1500   & 1500 & 831.8  \\
$m_{1/2}$   & 160    & 160  & 161.2  \\
$A_0$       &   0    &$-$900& $-$1597.1 \\
$\tan\beta$ & 54.8   & 52.1 & 17.6  \\
$m_t$       & 175    & 170.9  & 170.9 \\
$\mu$       & 203.5  & 177.5 & 644.0  \\
$m_{\tg}$   & 472.3  & 476.9 & 450.8  \\
$m_{\tu_L}$ & 1522.2 & 1522.8 & 891.1  \\
$m_{\tu_R}$ & 1526.0 & 1526.5 & 914.4  \\
$m_{\tst_1}$& 897.1  & 890.7  & 248.3  \\
$m_{\tb_1}$ & 1022.1 & 1025.0 & 632.2  \\
$m_{\te_L}$ & 1501.6 & 1501.6 & 853.6  \\
$m_{\te_R}$ & 1499.8 & 1499.8 & 802.4  \\
$m_{\tw_1}$ & 110.9  & 106.3 & 131.7  \\
$m_{\tz_2}$ & 110.4  & 106.7 & 131.0  \\ 
$m_{\tz_1}$ & 62.4   &  61.8 & 66.6  \\ 
$m_A$       & 309.1  & 347.0 & 157.0  \\
$m_h$       & 113.6  & 112.8 & 116.6   \\ \hline
$\Omega_{\tz_1}h^2$& 0.11  & 0.11 & 0.10   \\
$BF(b\to s\gamma)$ & $3.0\times 10^{-4}$ & $2.4\times 10^{-4}$ & 
$3.1\times 10^{-4}$ \\
$\Delta a_\mu    $ & $10.3 \times  10^{-10}$ & $10.0 \times  10^{-10}$ 
& $5.4 \times  10^{-10}$ \\
$BF(B_s\to\mu^+\mu^- )$ & $2.2\times 10^{-9}$ & $9.3\times 10^{-9}$ 
& $3.7\times 10^{-8}$  \\
$\sigma_{sc} (\tz_1p )\ [{\rm pb}]$ 
  & $3.2\times 10^{-7}$ 
  & $3.1\times 10^{-7}$ 
  & $2.6\times 10^{-8}$ \\
$\langle\sigma v\rangle |_{v\to 0}\ (cm^3/sec)$ & $2.0\times 10^{-26}$ & $ 2.3\times 10^{-26}$ &
$1.6\times 10^{-26}$ \\
\hline
\end{tabular}
\caption{Masses and parameters in~GeV units
for three Egret-motivated benchmark points using Isajet 7.76.
}
\label{tab:bm}
\end{table}

\begin{figure}[!t]
\begin{center}
\epsfig{file=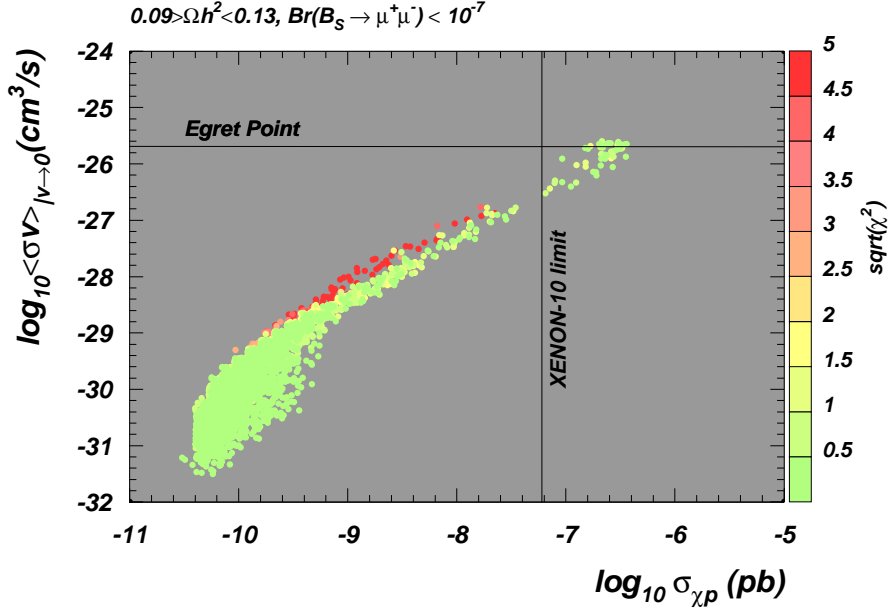,width=12cm}
\end{center}
\caption{\small\it 
Scan of mSUGRA parameter space for models which obey LEP2 constraints and have 
$0.09< \Omega_{\tz_1}h^2<0.13$.
We plot results in the $\langle\sigma v\rangle|_{v\to 0}\ vs.\ 
\sigma_{SI}(\tz_1 p)$ plane.
Green dots have good $BF(b\to s\gamma )$ while red dots 
deviate from the measured branching fraction.}
\label{fig:sug} 
\end{figure}

We next wish to check if Xenon-10 would exclude {\it all} mSUGRA interpretations of the
Egret data. We do so by scanning over the entire mSUGRA model parameter space:
\begin{eqnarray}
100~{\rm GeV} <       &m_0&	      <4000~{\rm GeV}\nonumber\\
10~{\rm GeV}  <       &m_{1/2}& 	      <1000~{\rm GeV}\nonumber\\
-3000~{\rm GeV}<      &A_0&	      <3000~{\rm GeV}\nonumber\\
1.1<		      &\tan\beta&     <60     \nonumber\\
        	      &\mu&>0 .
\label{eq:paraspace}
\end{eqnarray}
Our results of this scan are plotted in Fig. \ref{fig:sug}. Here, we keep only points 
with $0.09<\Omega_{\tz_1}h^2<0.13$, and also $50\ {\rm GeV}<m_{\tz_1}<70$ GeV.
Green points have a $BF(b\to s\gamma )$ in close accord with the measured value:
$BF(b\to s\gamma )=(3.55\pm 0.26)\times 10^{-4}$ from a combined analysis \cite{bsg_ex}
of the CLEO, Belle and BABAR experiments. Yellow and especially red points give 
branching fractions further from the experimental central value.
The surviving points are plotted in the $\langle\sigma v\rangle |_{v\to 0}\ vs.\ 
\sigma_{SI}(\tz_1 p)$ plane. We see that most points populate the low
 $\langle\sigma v\rangle |_{v\to 0}$ and low $\sigma_{SI}(\tz_1 p)$ regions.
These points come from either the stau co-annihilation region or the $h$-resonance
annihilation region of the mSUGRA model. 
There are no hyperbolic branch/focus point (HB/FP) contributions since $m_{\tz_1}<M_W$, and so
$\tz_1\tz_1\to W^+W^-$ (as is enhanced in the HB/FP region) is kinematically forbidden.
The points at large $\langle\sigma v\rangle |_{v\to 0}\sim 10^{-26}\ {\rm cm}^2/{\rm sec}$
(so that they have a high halo annihilation rate into gammas) also are {\it all above the 
Xenon-10 dark matter limit}!  For this reason, it seems an interpretation of the Egret GeV anomaly 
in terms of neutralino annihilation in the mSUGRA model is ruled out.
Of course, one way out is to assume we live in a local void of dark matter, and the local
density is far  below the assumed value of 0.3 GeV/cm$^3$. One may also assume much lower
local WIMP velocities, which would also lead to lower detection rates.
(It may also be the case that we live in a locally overdense region, or that the
velocity profile is harder than expected, leading to larger than expected DM detection
rates.)
Here, we will not further entertain these possibilities.

\subsection{NUHM2 analysis}

While the mSUGRA model does not seem adequate to explain the Egret GeV anomaly
in the face of the new Xenon-10 limit, other less restrictive supersymmetric models
may do the job. One highly motivated model beyond mSUGRA consists of models
with non-universal soft SUSY breaking Higgs masses (NUHM). In simple $SO(10)$ SUSYGUT
theories, the Higgs supermultiplets live in a {\bf 10} of $SO(10)$, while the
matter supermultiplets live in the {\bf 16} dimensional spinor representation. 
Thus, one might naturally expect Higgs SSB terms to have different GUT scale masses than matter 
SSB terms (this is the one-parameter NUHM model, or NUHM1\cite{nuhm1}). 
In $SU(5)$ SUSYGUT models, the doublet $\hat{H}_u$ lives in a {\bf 5}, 
while the doublet $\hat{H}_d$ lives in a $\overline{\bf 5}$. In this case, both
$m_{H_u}^2$ and $m_{H_d}^2$ can be taken as independent parameters, whereas the matter scalars
remain unified to $m_0$. This is the two-parameter NUHM model, or NUHM2\cite{nuhm2}.
In NUHM2, the GUT scale parameters $m_{H_u}^2$ and $m_{H_d}^2$ 
can be traded for independent weak scale parameters 
$\mu$ and $m_A$ (whereas in mSUGRA, these quantities are derived from the GUT scale
inputs, mainly $m_0$). Here, we will examine the NUHM2 model, with parameter
space given by
\be
m_0,\ m_{1/2},\ A_0,\ \mu ,\ m_A,\ \tan\beta ,
\ee
where we take $m_t=171$ GeV as usual.
Our goal will be to lower $\tan\beta$, so that we will diminish the
direct detection cross section to levels below the Xenon-10 limit.
Meanwhile, we wish to maintain a large $\langle\sigma v\rangle |_{v\to 0}$ so that
we maintain a high rate of neutralino halo annihilations. This can be done by 
lowering $m_A$ so that we move nearer (but not directly on) $A$-resonance
annihilation.

\begin{figure}[!t]
\begin{center}
\epsfig{file=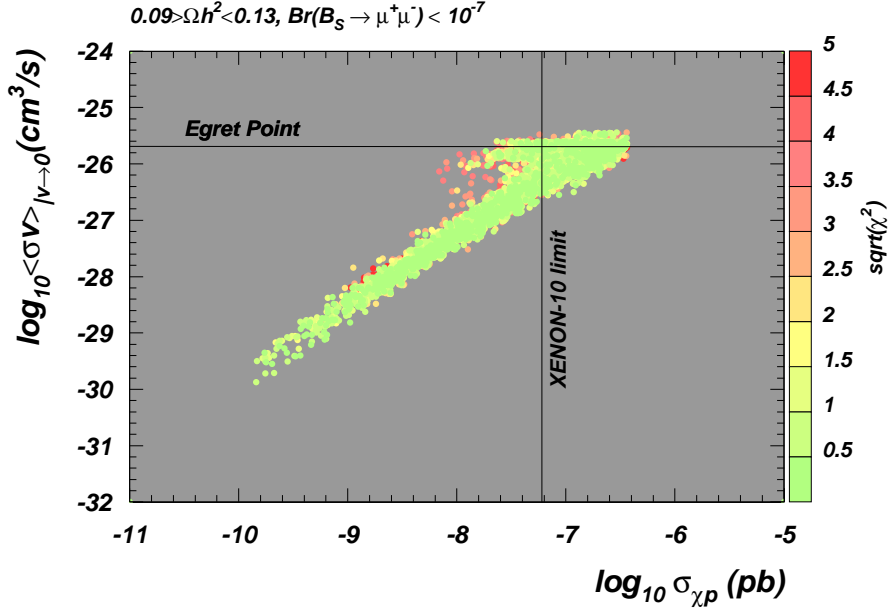,width=12cm}
\end{center}
\caption{\small\it 
Scan of NUHM2 parameter space for models which obey LEP2 constraints and have 
$0.09< \Omega_{\tz_1}h^2<0.13$.
We plot results in the $\langle\sigma v\rangle|_{v\to 0}\ vs.\ 
\sigma_{SI}(\tz_1 p)$ plane.
Green dots have good $BF(b\to s\gamma )$ while red dots 
deviate from the measured branching fraction.}
\label{fig:nuhm} 
\end{figure}

Here, we scan over the NUHM2 parameter space:
\begin{eqnarray}
100~{\rm GeV} <       &m_0&	      <2000~{\rm GeV}  \nonumber\\
50~{\rm GeV}  <       &m_{1/2}&        <300~{\rm GeV}   \nonumber\\
-3000~{\rm GeV}<      &A_0&	      <3000~{\rm GeV}  \nonumber\\
1.1<		    &\tan\beta&         <60            \nonumber\\
50~{\rm GeV}<         &\mu&             <1000~{\rm GeV}  \nonumber\\
70~{\rm GeV}<         &m_A&	      <500~{\rm GeV}	      ,
\label{eq:nuhm_pspace}
\end{eqnarray}
while again plotting points which satisfy the WMAP relic density bound
and have $50\ {\rm GeV}<m_{\tz_1}<70$ GeV. The points are again plotted in the
$\langle\sigma v\rangle |_{v\to 0}\ vs.\ \sigma_{SI}(\tz_1 p)$ plane in
Fig. \ref{fig:nuhm}. In this case, we {\it do} find a collection of points
with simultaneously the correct relic abundance and neutralino mass, a 
high rate of halo annihilation (since $\langle\sigma v\rangle |_{v\to 0}\sim 10^{-26}$cm$^3$/sec), 
{\it and} with $\sigma_{SI}(\tz_1 p)$ below the Xenon-10 limit!
While the collection of points does satisfy the Xenon-10 limit, note that they do not
extend to arbitrarily small values of $\sigma_{SI}(\tz_1 p)$. In fact, one prediction
is that if the NUHM2 SUGRA model is to explain the Egret GeV anomaly, then
$\sigma_{SI}(\tz_1 p)\agt 10^{-8}$ pb, which is well within range of a number
of upcoming direct detection experiments, including Lux, Xenon-100, WARP-140
and mini-CLEAN.

As an example of a NUHM2 point which satisfies all constraints, we list
the third point in Table \ref{tab:bm}: a point with $m_0=831.8$ GeV and
$m_{1/2}=161.2$ GeV. 
Like the mSUGRA models, it has a light gluino, with $m_{\tg}\sim 450$ GeV.
It also has $m_A\sim 157$ GeV, which is generically  below 
the mSUGRA models, which predict $m_A\sim 200-400$ GeV. 
The $\tan\beta$ value is just 17.6, so the $b$-Yukawa coupling is not so large, 
and the $A$-width is rather narrow.

\begin{figure}[h]
\includegraphics[width=0.55\textwidth]{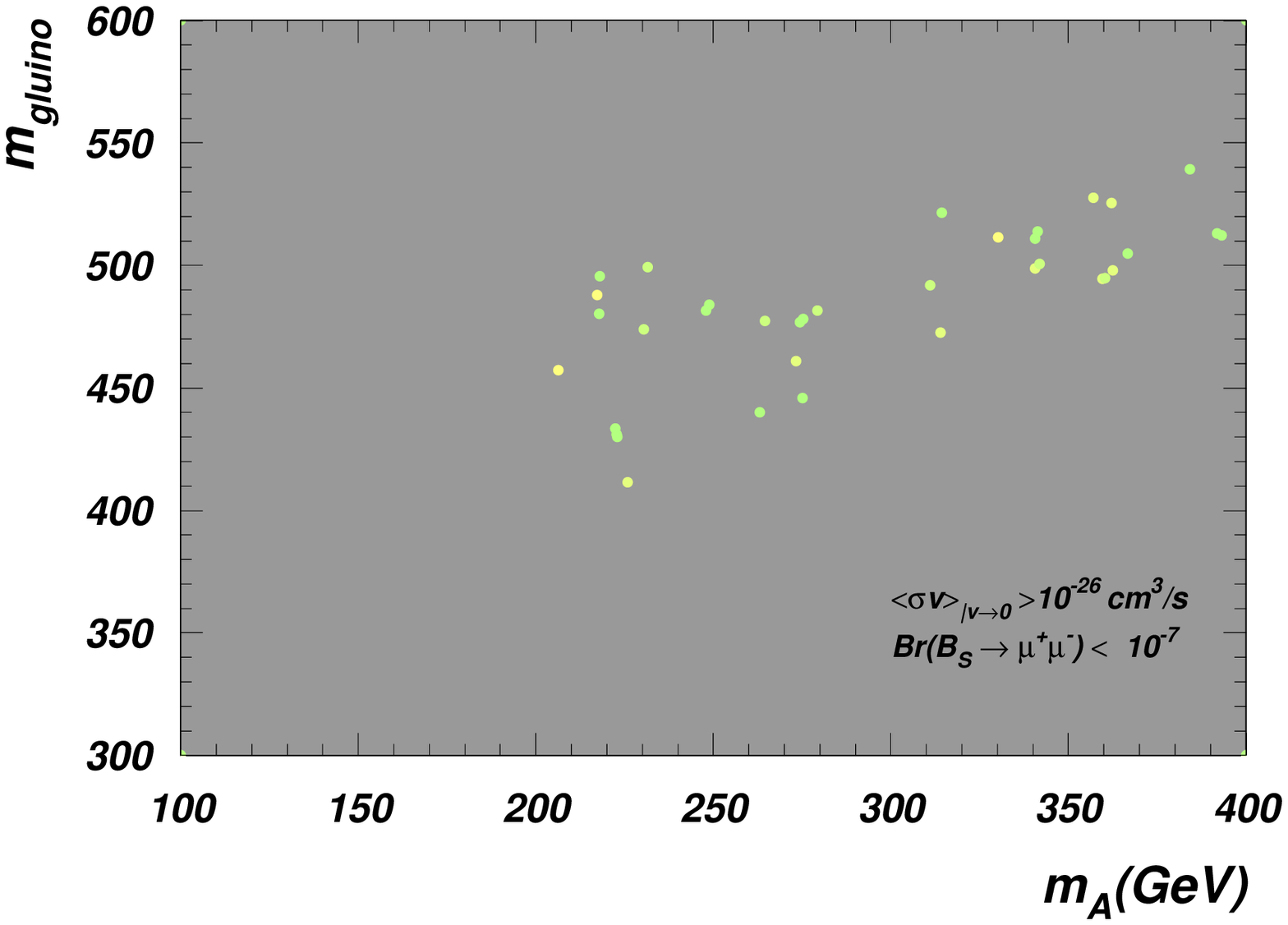}%
\hskip -1cm
\includegraphics[width=0.55\textwidth]{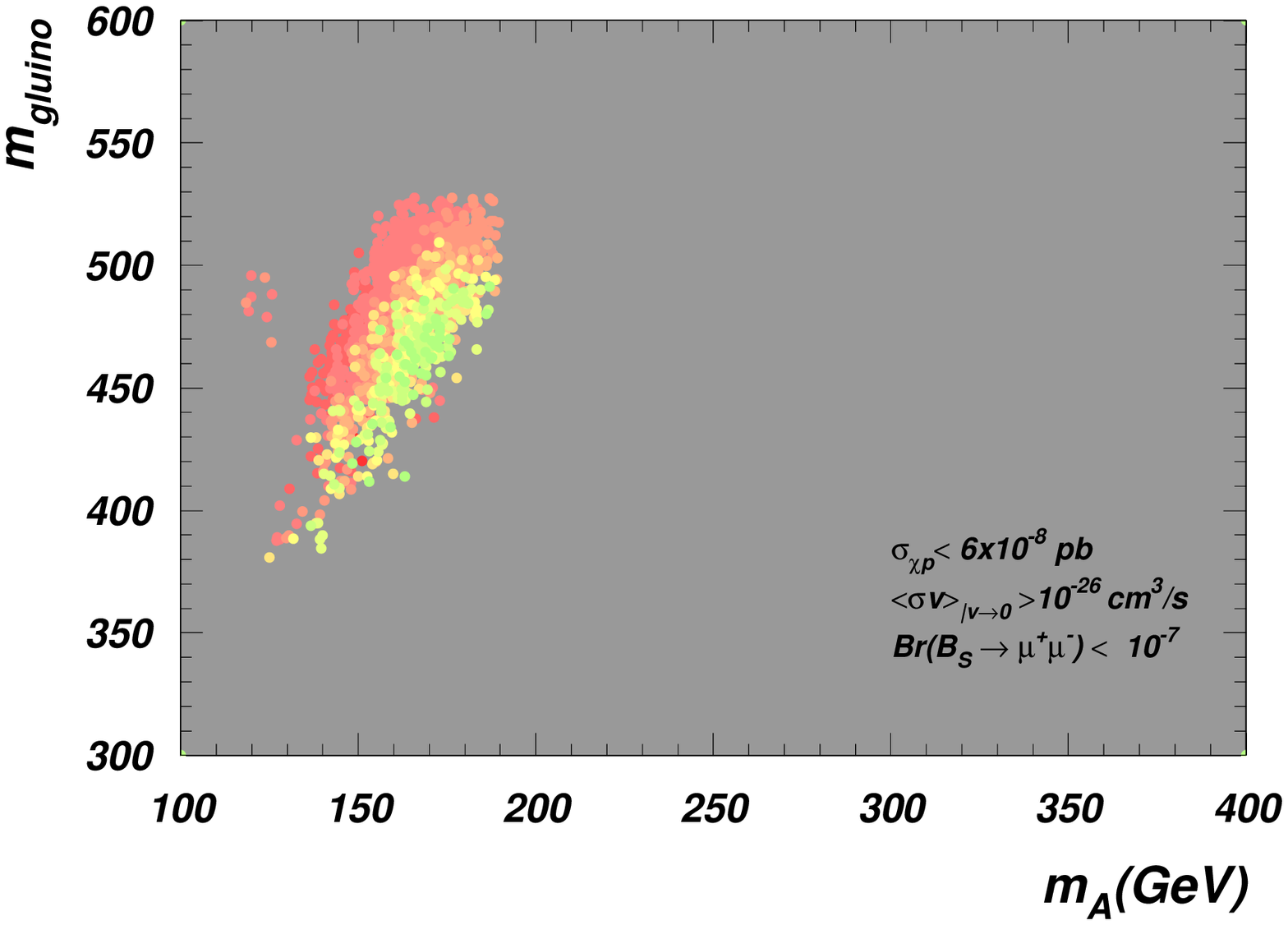}
\vskip -5.cm
\hspace*{0.5\textwidth}\hspace*{-7cm}{\bf (a)}
\hspace*{0.5\textwidth}\hspace*{-0.8cm}{\bf (b)}
\vskip 5.cm
\caption{\small\it 
Scan of {\it a}). mSUGRA model and {\it b}). NUHM2 model 
parameter space for points which obey LEP2 constraints and have 
$0.09< \Omega_{\tz_1}h^2<0.13$ and $\langle\sigma v\rangle|_{v\to 0}>10^{-26}$ cm$^3$/sec.
We plot results in the $m_{\tg}\ vs.\ m_A$ plane.
Green dots have good $BF(b\to s\gamma )$ while red dots 
deviate from the measured branching fraction.}
\label{fig:mglmA} 
\end{figure}

To show explicitly the range of $m_{\tg}$ and $m_A$ expected in 
our scans, we plot in Fig. \ref{fig:mglmA} points from {\it a}). the mSUGRA scan and
{\it b}). the NUHM2 scan in the $m_A\ vs.\ m_{\tg}$ plane.
In both cases, surviving points have $m_{\tz_1}:50-70$ GeV and $0.09<\Omega_{\tz_1}h^2<0.13$,
$BF(B_s\to\mu^+\mu^- )<10^{-7}$ and further, 
we require $\langle\sigma v\rangle|_{v\to 0}>10^{-26}$ cm$^3$/sec so that there
is sufficient halo annihilation to explain the Egret excess. The points in the mSUGRA case all have 
$m_{\tg}\sim 400-550$ GeV, while $m_A>200$ GeV. In the NUHM2 case, we further require
$\sigma_{SI}(\tz_1 p)<6\times 10^{-8}$ pb. Here, we see a similar range of $m_{\tg}$ is allowed.
However, in the NUHM2 case, we always have $m_A<200$ GeV. This is an important distinction 
between the two interpretations which can be directly tested/measured at LHC.

\section{Egret SUSY interpretation and the LHC}

The CERN LHC is expected to turn on in mid-2008, and gradually begin
accumulating data into 2009. It is not unreasonable to expect of order
0.1 fb$^{-1}$ of integrated luminosity in the first full year of running.

If the SUSY interpretation of the Egret GeV anomaly is correct, then
we expect the DM particle to be a neutralino of mass 
$m_{\tz_1}\simeq 50-70$ GeV. Most SUSY models also assume
gaugino mass unification at the GUT scale. In such models, the
gluino mass is expected to be $m_{\tg}\sim 7 M_1$, and if
the $\tz_1$ is dominantly bino-like, with $\mu \gg M_1$, then
we expect $m_{\tz_1}\sim M_1$ and thus would predict
$m_{\tg}\sim 350-500$ GeV. Squarks and sleptons can be considerably heavier, 
as is the case in the points listed in Table \ref{tab:bm}.
In this type of scenario, we would expect new physics events from SUSY 
at the CERN LHC to be dominated by gluino pair production, followed by 
gluino cascade decays\cite{cascade}. 
The cross section for $pp\to\tg\tg X$ at the LHC 
for the above range of gluino masses is $\sim 10^4-10^5$ fb\cite{sig_gg}. 
Thus, with just 0.1 fb$^{-1}$ of integrated luminosity, we can already expect
$10^3-10^4$ gluino pair events to be recorded per LHC experiment.

While new physics may be lurking in the LHC already shortly after
turn-on, it is unclear whether the detectors will be fully calibrated
to allow for a new physics search. For instance, traditional SUSY searches
rely on a $\eslt +jets$ signature, where large $\eslt$
is required to reject SM background events from the SUSY signal.
To properly use the $\eslt$ variable, a full knowledge of the detector is
required, since $\eslt$ can not only arise from signal and background 
events, but also from 1. dead regions of the detector, 2. ``hot'', or
mis-firing calorimeter cells, 3. cosmic ray events and 4.
energy mis-measurement in active calorimeter cells.
In Ref. \cite{etmiss}, it was recently pointed out that early 
discovery of SUSY at the LHC was possible {\it without} using 
$\eslt$, and that a reach in $m_{\tg}$ to $600-700$ GeV
could be attained without using $\eslt$, and with only 0.1 fb$^{-1}$
of integrated luminosity. Effectively, the idea was to make use of
multi-lepton production\cite{multilep} in the lengthy sparticle cascade decays.
Thus, requiring $\ge 4$ jets production along with $\ge 2$ or $3$
isolated leptons, allowed for an excellent rejection of SM background
(dominated by $t\bar{t}$ production) compared to signal so that
a SUSY discovery could be made. The values of $m_{\tg}$ expected from
the SUSY interpretation of the Egret GeV anomaly fall well within
this ``early discovery'' range.  

\begin{figure}[!t]
\begin{center}
\epsfig{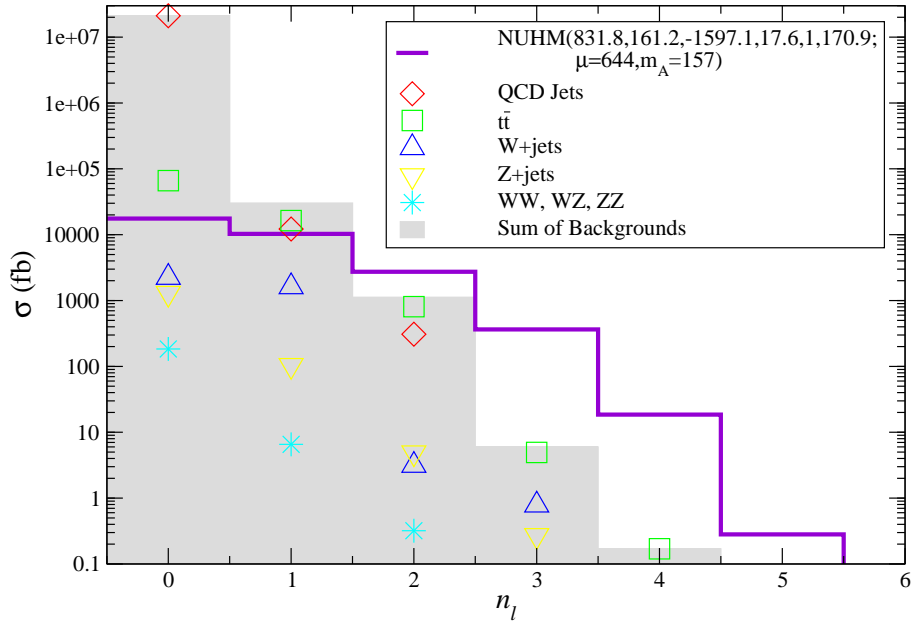}
\end{center}
\caption{\small\it 
Distribution in number of isolated leptons from the NUHM2
point in Table \ref{tab:bm} after cuts $C1^\prime$. We also show
component SM backgrounds and the summer SM background (gray histogram).
}
\label{fig:nl} 
\end{figure}

Here, we use the same detector simulation, jet finding algorithm and lepton
isolation criterion as 
detailed in Ref. \cite{etmiss}, and adopt the same set of SM background
events from QCD jet production, $W+jets$, $Z+jets$, $t\bar{t}$ production
and vector boson pair production. We require first that signal and
background events satisfy the set of cuts $C1^\prime$:
{\it i}). $n(jets)\ge 4$, {\it ii}). $E_T(j1,j2,j3,j4)>100,\ 50,\ 50,\ 50$ GeV, 
respectively (where jets are ordered according to $E_T$ value) and
{\it iii}). transverse sphericity $S_T\ge 0.2$. 
We then plot a multiplicity of isolated leptons (a lepton $\ell =e$ or $\mu$ is isolated if
it has $E_T(\ell)>20$ GeV, $|\eta(\ell)|<2.5$ and $\sum E_T^{cells}<5$ GeV in a cone
of $\Delta R\equiv \sqrt{\Delta\eta^2 +\Delta\phi^2}<0.2$ about the lepton direction).
The results are shown in Fig. \ref{fig:nl}. 
Here, we see that SM background dominates signal for $n_\ell =0$ or 1. But already 
at $n_\ell =2$, the signal from the NUHM2 point in Table \ref{tab:bm} stands out above 
background. At $n_\ell =3$, there still remains $\sim 400$ fb of signal, while BG is negligible.
Given these results, it seems that the SUSY interpretation of the Egret GeV anomaly
should be easily testable at the LHC after only 0.1 fb$^{-1}$ of integrated 
luminosity is obtained, and even before the detectors are fully calibrated such that 
they can perform a reliable search for $\eslt +jets$ events.

\begin{figure}[!t]
\begin{center}
\epsfig{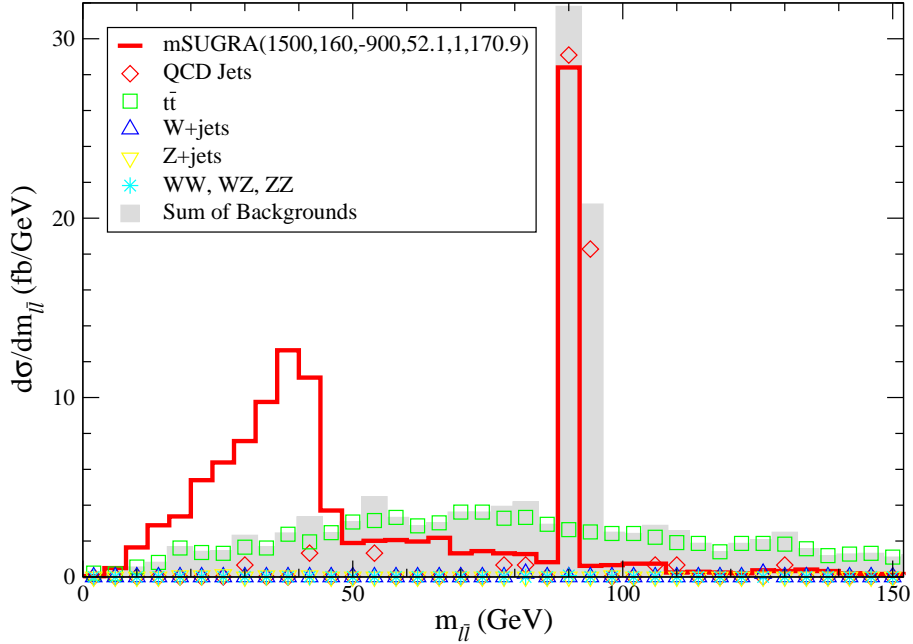}
\end{center}
\vspace*{-0.5cm}
\caption{\small\it 
Distribution in OS/SF dilepton invariant mass from the second mSUGRA
point in Table \ref{tab:bm} after cuts $C1^\prime$. We also show
component SM backgrounds and the sum of SM background (gray histogram).
}
\label{fig:mll} 
\end{figure}

Once a SUSY signal is obtained, then the set of likely signal events can be scrutinized
to try to reconstruct sparticle masses, etc. The starting point is often to first look
at the opposite sign/same flavor (OS/SF) dilepton invariant mass spectrum $m(\ell\bar{\ell})$\cite{mll}.
We plot the distribution $d\sigma /dm(\ell\bar{\ell})$ in Fig. \ref{fig:mll} arising
from the second mSUGRA point in Table \ref{tab:bm}, after cuts $C1^\prime$ and requiring
a pair of OS/SF isolated leptons.
In this case, a clear mass edge is seen at $m_{\tz_2}-m_{\tz_1}=44.9$ GeV. There is also
a $Z$ peak in both signal and BG (the latter arising because Isajet includes $W$ and $Z$
radiation in its parton shower algorithm).

\begin{figure}[!t]
\begin{center}
\vspace*{1cm}
\epsfig{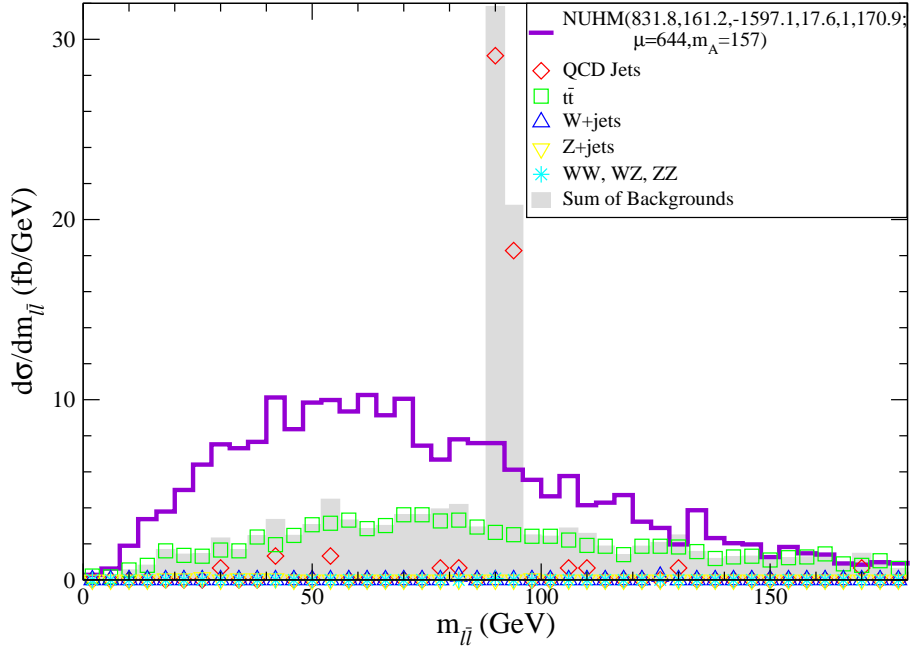}
\end{center}
\vspace*{-0.5cm}
\caption{\small\it 
Distribution in OS/SF dilepton invariant mass from the NUHM2
point in Table \ref{tab:bm} after cuts $C1^\prime$. We also show
component SM backgrounds and the sum of SM background (gray histogram).
}
\label{fig:mll_nuhm} 
\end{figure}

\begin{figure}[tbp]
\includegraphics[width=0.55\textwidth]{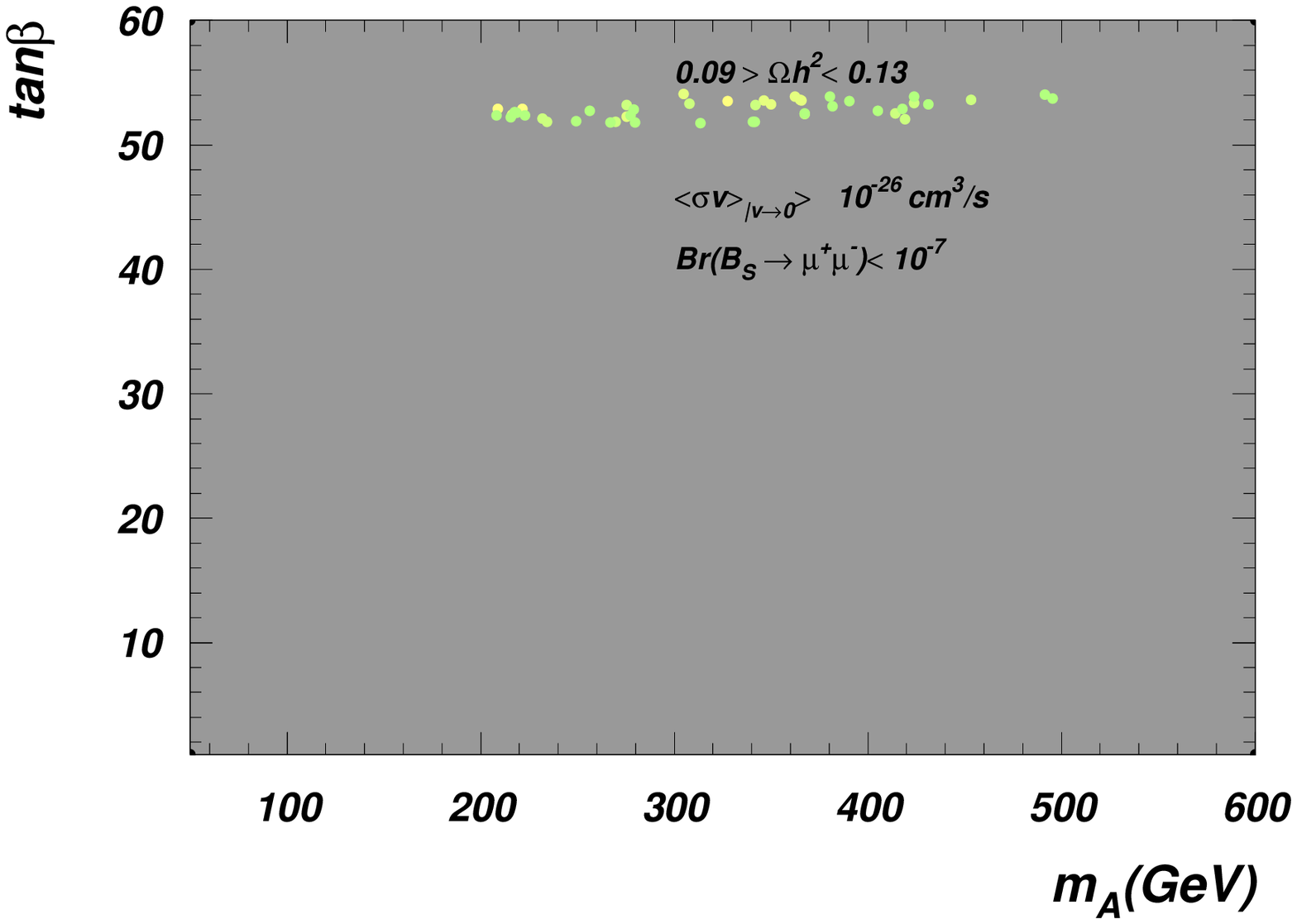}%
\hskip -1cm
\includegraphics[width=0.55\textwidth]{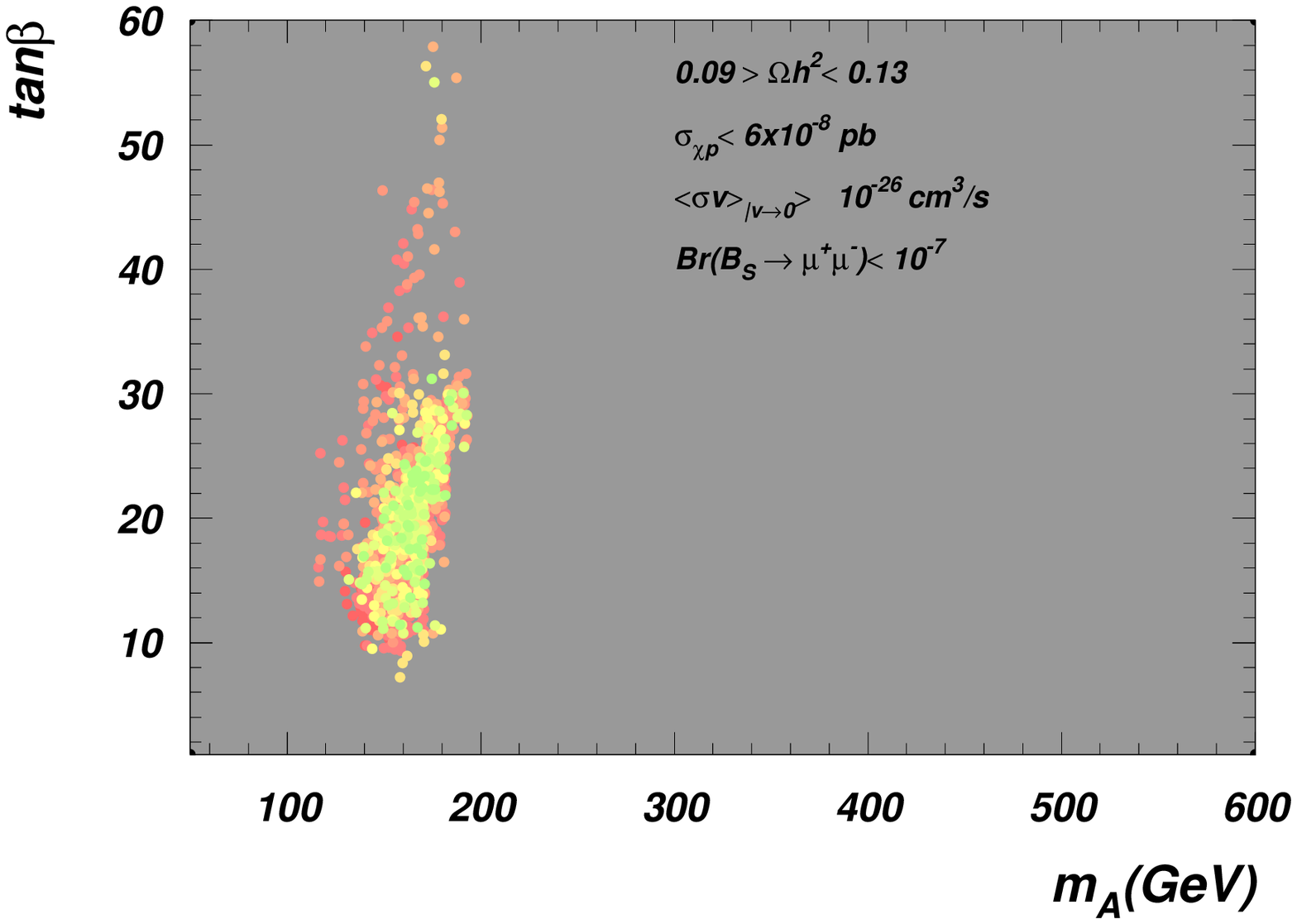}%
\vskip -5.cm
\hspace*{0.5\textwidth}\hspace*{-7cm}{\bf (a)}
\hspace*{0.5\textwidth}\hspace*{-0.8cm}{\bf (b)}
\vskip 4cm
\caption{\small\it 
Points from {\it a}). mSUGRA  and {\it b}).  
NUMH2 interpretation of the Egret GeV anomaly in $(\tan\beta,\ m_A)$ plane
with lightest  neutralino mass in the range  of 50-70 GeV.
Points  obey LEP2 constraints and have 
$0.09< \Omega_{\tz_1}h^2<0.13$,
$\langle\sigma v\rangle|_{v\to 0}>10^{-26}$ cm$^3$/sec
and 
$Br(B_S\to \mu^+\mu^-)<10^{-7}$.
For NUMH2 parameter space we also require $\sigma_{SI}(\tz_1 p)<6\times 10^{-8}$~pb
to satisfy Xenon-10 limit.
Green dots have good $BF(b\to s\gamma )$ while red dots 
deviate from the measured branching fraction.
\label{tanb_vs_ma}}
\end{figure}

In the case of the NUHM2 model from Table \ref{tab:bm}, the $\tst_1$ mass is so light
that $\tg\to t\tst_1$ dominantly, and the $\tz_2$ production via cascade decays is somewhat 
suppressed. Furthermore, the $\tz_2\to\tz_1 e^+e^-$ branching fraction is suppressed to the 0.8\% level;
in this case, the suppression is due to the presence of relatively light $A$ and $H$ Higgs bosons, 
which enhance the
decay $\tz_2\to\tz_1b\bar{b}$ to the 45\% level, at the expense of first/second generation
decay modes.
Thus, in the $m(\ell\bar{\ell})$ distribution for NUHM2 shown in Fig. \ref{fig:mll_nuhm},
we see a continuum distribution instead of a distinct mass edge (the mass edge would occur 
at $m_{\tz_2}-m_{\tz_1}=64.4$ GeV in this case). 
The other crucial observable to test between a mSUGRA or NUHM2 interpretation of the Egret excess
will come from the LHC measurement of the heavy Higgs boson mass spectrum.
The Atlas and CMS groups have posted reach plots for MSSM Higgs bosons in the $m_A\ vs.\ \tan\beta$
plane\cite{atlas,cms}. The $H$ and $A$ Higgs bosons should be seeable in the $b\bar{b}$, $\tau\bar{\tau}$ and
even $\mu^+\mu^-$ modes\cite{chung} in the NUHM2 model for $\tan\beta \agt 6$ since $m_A\alt 200$ GeV,
and their masses should be measureable. The mSUGRA interpretation requires $\tan\beta \sim 50$ and has 
relatively light $m_A$ as well, and should likewise be visible, but with mass $m_A\agt 200$ GeV.
In Fig.~\ref{tanb_vs_ma}, we present the mSUGRA {\it a}). and  NUHM2 {\it b}).
points in the $(\tan\beta,\ m_A)$ plane with lightest  neutralino mass in the range  of 50-70 GeV.
Points  obey LEP2 constraints and have 
$0.09< \Omega_{\tz_1}h^2<0.13$,
$\langle\sigma v\rangle|_{v\to 0}>10^{-26}$ cm$^3$/sec
and 
$Br(B_S\to \mu^+\mu^-)<10^{-7}$.
For NUMH2 parameter space we also require $\sigma_{SI}(\tz_1 p)<6\time 10^{-8}$~pb
to satisfy the Xenon-10 limit.
One can clearly see that indeed these scenarios suggest a very distinctive $A$-Higgs
boson phenomenology. Recent Tevatron limits~\cite{aspen_higgs} for the $(\tan\beta,\ m_A)$ plane
already exclude $m_A<180$ for $\tan\beta\ge 50$. So, the Tevatron can test now a
small part of NUMH2 parameter space with large values of  $\tan\beta$, while 
parameter space with $100<m_A<200$ and  $\tan\beta\le 30$ could be tested
and completely covered only at the LHC.

\section{Conclusions}

In this paper, we have examined the SUSY interpretation of the Egret GeV anomaly.
The SUSY interpretation requires $m_{\tz_1}\sim 50-70$ GeV, and a relic density of
$\Omega_{\tz_1}h^2\sim 0.1$. In order to satisfy these criteria, while maintaining a large
halo annihilation rate for $\tz_1\tz_1\to A^*\to b\bar{b}$, one must move to
very large $\tan\beta\sim 52-55$ in the context of the mSUGRA model.
At this high $\tan\beta$, the predicted spin-independent $\tz_1 p$ scattering
cross section exceeds recent limits from the Xenon-10 collaboration.

In order to maintain a SUSY interpretation of the Egret GeV anomaly, we suggest moving to
SUSY models with a non-universal Higgs sector: the NUHM2 model. In this case, with
freedom to adjust the value of $m_A$, one can reduce the value of $\tan\beta$
so that the predicted $\sigma_{SI}(\tz_1 p)$ is lowered below Xenon-10 limits, but maintain
a valid relic density and halo annihilation rate by lowering $m_A$ to values below $200$ GeV.
If the SUSY interpretation of the Egret GeV anomaly is correct, then we predict
direct detection cross sections $\sigma_{SI}(\tz_1 p)\agt 10^{-8}$ pb, which should be accessible 
to the next round of DD experiments. Further, the gluino mass should be in the range $400-550$ GeV,
and should be seeable in the multi-jet plus multi-lepton channel at LHC with just 0.1 fb$^{-1}$
of integrated luminosity. A dilepton mass edge in these signal events may not be apparent since the light
Higgs spectrum enhances $\tz_2$ decay to third generation fermions at the expense of first/second
generation leptons. Since $m_{A,H}\alt 200$ GeV, the $A$ and $H$ should be readily visible at LHC via 
searches for $A,\ H\to b\bar{b},\ \tau^+\tau^-$ or $\mu^+\mu^-$, even for relatively low values of
$\tan\beta\agt 6-10$. This observation clearly distinguishes between the mSUGRA and NUHM2 
interpretations of the Egret GeV anomaly.

\section*{Acknowledgments} We gratefully acknowledge 
Professor W. de Boer for various discussions. 
%

\end{document}